\documentclass[num-refs]{wiley-article}

\usepackage{siunitx}
\usepackage{multirow}
\usepackage{tcolorbox}
\usepackage{booktabs}
\usepackage{array}

\papertype{Empirical Article}
\paperfield{Journal Section}

\title{Empirical Evaluation of ChatGPT on Requirements Information Retrieval Under Zero-Shot Setting}

\abbrevs{IR, information retrieval; NLP, natural language procssing; RE, requirements engineering; NLP4RE, natural language procssing for requirements engineering; LLM, large language model; SE, software enginering.}

\author[1]{Jianzhang Zhang}
\author[1]{Yiyang Chen}
\author[2]{Nan Niu}
\author[3]{Yinglin Wang}
\author[1]{Chuang Liu}

\affil[1]{Department of Management Science and Engineering, Hangzhou Normal University, Hangzhou, Zhejiang, 311121, P.R.China}
\affil[2]{Department of Electrical Engineering and Computer Science, University of Cincinnati, Cincinnati, Ohio, 45221, USA}
\affil[3]{Department of Computer Science and Technology, Shanghai University of Finance and Economics, Shanghai, 200433, P.R.China}

\corraddress{Jianzhang Zhang, Department of Management Science and Engineering, Hangzhou Normal University, Hangzhou, Zhejiang, 311121, P.R.China}
\corremail{zjzhang@hznu.edu.cn}

\fundinginfo{This paper is supported by the National Natural Science Foundation of China (Grant No.61873080) and Engineering Research Center of Mobile Health Management System of Chinese Ministry of Education (Grant No. 2022GCZXGH04).}

\runningauthor{Zhang et al.}

\begin{document}

\begin{frontmatter}
\maketitle

\begin{abstract}
Recently, various illustrative examples have shown the impressive ability of generative large language models (LLMs) to perform NLP related tasks. ChatGPT undoubtedly is the most representative model. We empirically evaluate ChatGPT's performance on requirements information retrieval (IR) tasks to derive insights into designing or developing more effective requirements retrieval methods or tools based on generative LLMs. We design an evaluation framework considering four different combinations of two popular IR tasks and two common artifact types. Under zero-shot setting, evaluation results reveal ChatGPT's promising ability to retrieve requirements relevant information (high recall) and limited ability to retrieve more specific requirements information (low precision). Our evaluation of ChatGPT on requirements IR under zero-shot setting provides preliminary evidence for designing or developing more effective requirements IR methods or tools based on LLMs.

% Please include a maximum of seven keywords
\keywords{ChatGPT, Requirements Information Retrieval, Empirical Evaluation, Natural Language Processing for Requirements Engineering}
\end{abstract}
\end{frontmatter}

\section{Introduction}\label{sec1}

% one of critical success factors for  
% 满足用户需求既是产品成功的关键因素，也是软件项目开发活动的关键因素
% 泛在的软件系统成为人机物三元融合的基础设施之一，one of the fundamental \cite{mei2017understanding}
% 软件开发声明周期包括，需求，设计，编码，测试等
% 其中需求工程是什么，对软件开发至关重要\\
% 需求工程对于软件开发很重要
Ubiquitous software systems are acting as the fundamental infrastructure in the era of digital economy~\cite{wang2021managing}. Software-defined everything has been a broader trend in the information technology community~\cite{mei2017understanding}. Among the software lifecycle phases, requirements engineering is the process of eliciting individual stakeholder requirements and needs and developing them into detailed, agreed requirements documented and specified in such a way that they can serve as the basis for all other system development activities~\cite{pohl2010requirements}. The requirements engineering process plays a significant role on successfully evolving software projects and keeping competitive in the market~\cite{aurum2005engineering}. 

% 需求是什么
% 收集和分析用户需求很重要
% 传统的用户收集渠道有什么，过去十年随着App分发平台的涌现，社交媒体等成为新的渠道channels

% Anas
% 也会从App descriptions中挖掘feature 做推荐，TOSEM那篇和inspired那篇
% through software products platforms, e.g., Google Play and App Store, or social media, e.g., Twitter.
Requirements are a verbalization of decision alternatives regarding a system’s functionality and quality~\cite{aurum2003fundamental}. Traditional requirements elicitation usually involves users through surveys, interviews, workshops, and focus groups~\cite{maalej2015toward}. In the last decade, online platforms, e.g., Google Play and App Store~\cite{pagano2013user}, or social media, e.g., Twitter~\cite{williams2017mining}, have been serving as prevalent channels for eliciting user requirements. Feature request~\cite{malgaonkar2022prioritizing}, bug reports~\cite{gao2022understanding, gao2021emerging}, and non-functional concerns~\cite{jha2019mining} are common requirements information among others that are mined from user generated content such as App reviews and twitter posts. 
% feature request, bug report, and enhancement information
% non-functional information
Besides of user reviews in App sores, App descriptions are also treated as a valuable domain knowledge source for inspiring feature related requirements elicitation and recommendation~\cite{ferrari2023strategies, jiang2019recommending, ebrahimi2021classifying}.

% 需求主要是由文本、自然语言组成，自然语言是一种很重要的制品媒介，找几篇20年前后稍微新的文献，如ICSE等
%The important role of natural language (NL) in requirements engineering (RE) has long been established~\cite{ryan1993role}
% 随着人工智能的发展，ML/NLP等技术开始广泛应用于提高软件开发各阶段活动的效率，列举几篇做相关任务的ICSE论文
The rapid advancement of deep learning methods and techniques leads to the performance breakthroughs on many computer vision and natural language processing (NLP) tasks~\cite{lecun2015deep}. Meanwhile, a wide range of automated methods based on machine learning have been proposed to increase the effectiveness of software engineering lifecycle activities~\cite{kotti2023machine}. As natural language  are generally used to express software requirements~\cite{ferrari2017natural, ferrari2018natural} and its role in requirements engineering has long been established~\cite{ryan1993role}, applying natural language processing for requirements engineering (NLP4RE)~\cite{zhao2021natural} has drawn much attention from software engineering research community. Especially, LLMs have achieved new state of the arts in various NLP tasks in the last decade, which motivate SE researchers and practioners to propose and develop more effective methods and tools for supporting RE processing~\cite{luo2022prcbert,amaral2023nlp}.

% 上面从软工-需求工程-需求文本
% 下面从LLM到chat-NLP4RE

% chatGPT是代表，备受关注

% Therefore, the end-to-end multiple tasks performing advantage of ChatGP motivate us to explore its potential in requirements processing for RE. Specifically, our research goal in this paper is to quantitatively evaluate the effectiveness of ChatGPT in requirements information retrieval (IR). 

%  IR is one of the dominant techniques to support RE process~\cite{zhao2021natural}.
Generative LLMs have impressed research and industry communities with its capability to perform various NLP related tasks following human prompts in an chat-based and end-to-end manner~\cite{zhao2023survey}. The release of ChatGPT (Chat Generative Pre-trained) has draw wide attention due to its potentially profound impact on various domains1 . ChatGPT is driven by a generative large language model trained with reinforcement learning from human feedback~\footnote{https://openai.com/chatgpt}\label{blog}. Among others, natural language processing (NLP) and software programming are two prominent impacted domains, which can also glimpsed from the highlighted examples in ChatGPT's official blog\footref{blog}. This mainly stems from the code pretraining on large scale corpora and instruction tuning by human prompts~\cite{wu2023brief, zhou2023chatgpt}.

On the one hand, natural language requirements artifacts, including specialized (target to developers) and general ones (target to developers and users), spread all over different software repositories and RE phases~\cite{vidoni2022systematic}. For application in practice, additional efforts, e.g. implementation, deployment, and learning to use, are still needed. On the other hand, existing NLP4RE methods and tools are mostly designed for specific tasks to support certain RE phases~\cite{zhao2021natural}. With appropriate prompts, ChatGPT can perform almost all NLP tasks from a wide range of domains in a conversational manner, where the prompts can be viewed as human queries in a dialogue. 

These facts motivate us to evaluate the effectiveness of ChatGPT in performing different requirements IR tasks to derive insights into more effective requirements IR methods and tools. ChatGPT is the most representative generative LLMs~\cite{zhou2023chatgpt} and Requirements IR is the mostly performed NLP4RE task to support the RE process~\cite{zhao2021natural}, especially in the elicitation and analysis phases~\cite{wang2021}.

To this end, we empirically evaluate ChatGPT on 4 requirements benchmark datasets covering two popular IR tasks and two common artifact types under a zero-shot setting. Specific requirements IR tasks consist of requirements classification and feature extraction. Requirements artifacts include software requirements specification statements, App descriptions, and App reviews. The zero-shot setting could provide a performance lower bound.

In summary, the main contributions of this work are threefold:

%We make all of the evaluation materials, including data ,codes ,and results, publicly available\footnote{https://github.com/zhangjianzhang/ChatGPT4REIR\_eval} to facilitate reproduction and further research.
%
%The evaluation results provide preliminary evidence for designing or developing more effective requirements IR methods or tools based on generative LLMs. % 复制过来significant implication这句话

\begin{itemize}
	\item We design a framework for evaluating end-to-end chat-based LLMs on performing requirements IR tasks. Specifically, we empirically evaluate ChatGPT on retrieving requirements information from specialized and general artifacts. all of the evaluation materials are publicly available~\footnote{https://github.com/zhangjianzhang/ChatGPT4REIR\_eval}.
	
	 %具体测试了ChatGPT，代表性模型。也可是我们评测了，然后这个框架也通用于评测其他模型
	
	\item The quantitative and qualitative results demonstrates the feasibility and promising potential of employing LLMs to perform various requirements information retrieval tasks in an end-to-end manner.
	
	% reproduction和further research
	% 我们提供了一些建议
	\item we suggest possible directions of future efforts on the NLP4RE research and practice based on our evaluation results and the advancement of LLMs.
	
\end{itemize}

The rest of this paper is structured as follows: Section II details the proposed evaluation framework. Section III presents and analyzes the evaluation results quantitatively and qualitatively as well as threats to validity. Section IV discusses the implications of our evaluation providing insights into designing or developing more effective requirements retrieval methods or tools based on LLMs.  Section V presents the related work. In Section VI, we conclude the paper with future work.

\section{Evaluation Framework}\label{sec2}

\begin{figure}[t]
	\centerline{\includegraphics[width=\linewidth]{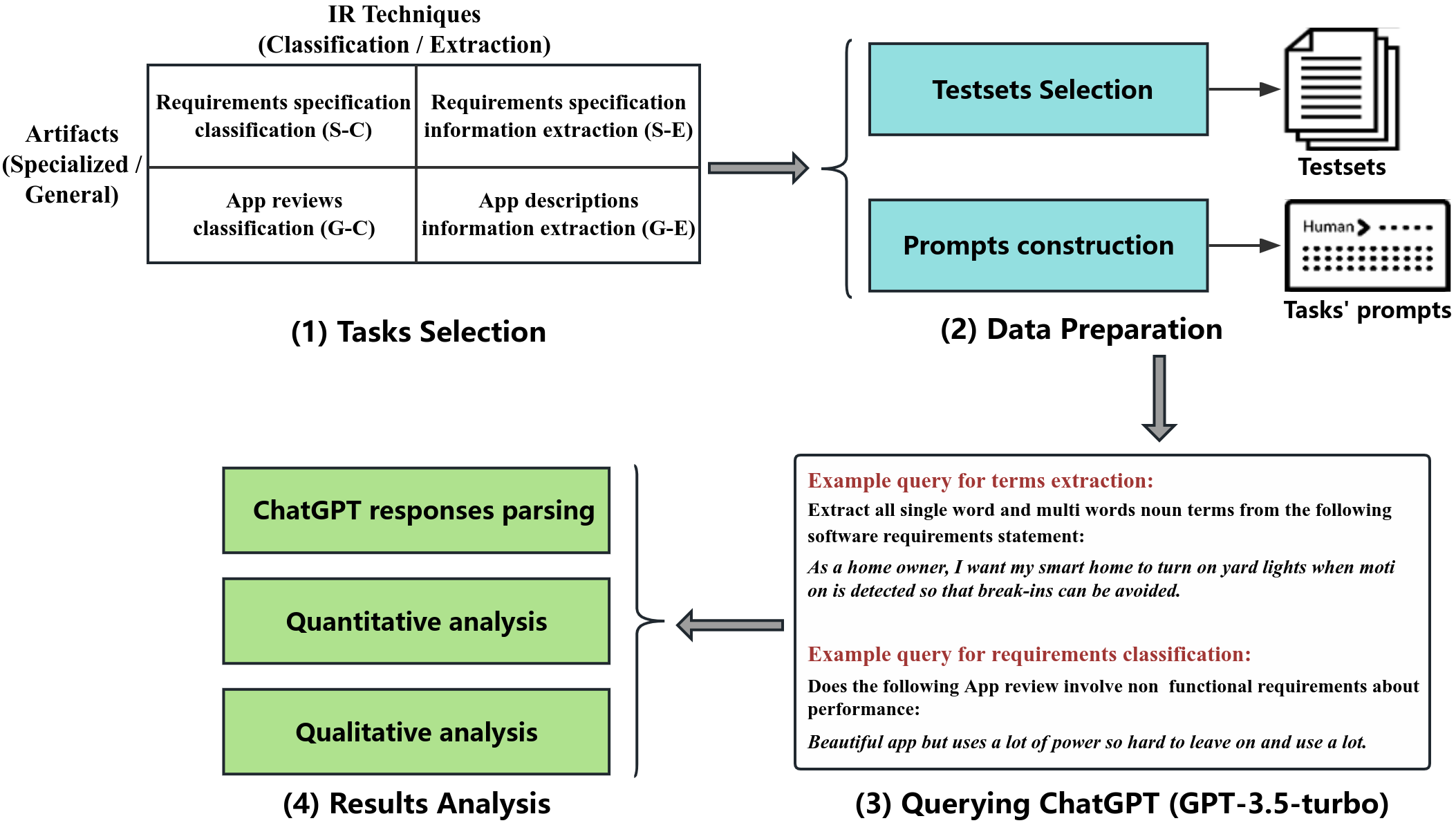}}\
	\caption{The Framework for Evaluating ChatGPT on Requirements IR Tasks}
	\label{fig:procedure}
\end{figure}

Figure \ref{fig:procedure} depicts our proposed evaluation framework. The whole framework consists of four sequential components: tasks selection, data preparation, querying ChatGPT, and results analysis. We elaborate each component of the framework in the following subsections. 

\subsection{Tasks Selection}

Evaluation tasks are formulated based on two dimensions of requirements IR. Specifically, we characterize the tasks with the involved IR techniques and RE artifact types. 

For the IR techniques, we select classification and extraction as both of them are most commonly employed techniques to facilitate the RE process. According to the recent survey~\cite{zhao2021natural}, more than 35\% (36.76\%) of NLP4RE studies utilize classification and extraction as their main techniques. In the context of requirements IR, the meaning of classification is to classify requirements into different categories, e.g., functional and quality categories. The extraction aims to identify key domain abstractions and concepts, such as, domain terms and feature-describing phrases.

For for the artifact types, we select specialized requirements specifications and general natural language documents as they are most dominant input document types for NLP4RE research~\cite{zhao2021natural, wang2021}. We refer the specialized requirements specifications to those textual requirements artifacts that are mainly target to developers, e.g., software requirements specifications (SRS) and user stories. In contrast, general natural language documents are targeted to both developers and users, e.g., App descriptions and reviews.

Combining the above two dimensions, a $2 \times 2$ evaluation task matrix can be formulated as shown in the first part of Figure ~\ref{fig:procedure}. The horizontal axis denotes IR techniques and the vertical axis denotes artifact types. For instance, the acronym $S\mbox{-}E$ means extracting information from specialized requirements specifications. In our evaluation, we select SRS statements and user stories to represent specialized requirements specifications. For general natural language documents, we select App descriptions and App reviews. The rationale behind of that selection is that those four requirements artifacts are commonly used in NLP4RE research~\cite{zhao2021natural, malgaonkar2022prioritizing, ferrari2023strategies}.

% 横轴表示什么，纵轴表示什么。
% (i.e., software requirements specifications, user stories) and general natural language documents (i.e., App descriptions and reviews), due to their prevalence in existing research and practice [1, 4, 5]. Finally, combining the above two dimensions, a $2 \times 2$ evaluation matrix is formulated as shown in the first part of Figure 1.

\subsection{Data Preparation}

% leading publication venues 
% 引用几篇综述说明数据集的来源是主要的SE研究出版源头
% Table \ref{tbl:data} shows data preparation results including introduction and descriptive text statistics of testsets and tasks' prompts.

Data preparation consists of testsets selection and prompts construction. Testsets selection aims to determine the dataset used to test the requirements IR ability of ChatGPT. Prompts construction is to craft the natural language queries to ChatGPT in order to instruct ChatGPT perform specific tasks on the selected dataset. Table \ref{tbl:data} summarizes the data preparation result including the brief introduction and descriptive text statistics of each dataset. We present the process and results of testsets selection and prompts construction in the remaining part of this subsection.

\begin{table*}[bt]
	\centering
	\caption{Evaluation testsets and tasks' prompts}
	
	\begin{tabular}{m{3.5cm}|m{5.1cm}|c|c|c|c}
		\toprule
		\multirow{1}{=}{\centering \textbf{Tasks}} & \multirow{1}{=}{\centering \textbf{Testsets Introduction}} & Count & Avg. Len. &  TTR & LD\\
		
		\midrule
		
		\multirow{1}{=}{\centering Requirements statements NFR multi-class classification (S-C)} & 249 non-functional requirements across 15 projects from the PROMISE NFR dataset. Each requirement statement is labeled with one of 4 NFR classes, i.e. \textit{usability, security, operational, and performance}. (\cite{hey2020norbert}, 2020) & 249 & 23.20 & 0.27 & 0.44\\
		\hline
		
		\multirow{1}{=}{\centering App review NFR multi-Label classification (G-C)} & 1800 reviews across various categories from Google Play and Apple App Store. Each review is assigned at least one NFR label from {\textit{dependability, performance, usability, supportability}} or a \textit{miscellaneous} label denoting the review not mentioning NFR. (\cite{jha2019mining}, 2019) & 1800 &33.33 & 0.19 & 0.43\\
		\hline
		
		\multirow{1}{=}{\centering Term extraction from user stories (S-E)} &  100 smarthome user stories. Each user story is stated with the template ``\textit{As a <Role>, I want my smart home to <Feature>, so that <Benefit>}''. A set of 250 domain terms are manually extracted from those requirements. (\cite{zhang2022automatic}, 2022) & 100 & 32.38 & 0.38 & 0.44\\
		\hline
		
		\multirow{1}{=}{\centering Feature extraction from App descriptions (G-E)} & 50 App Descriptions across 10 application categories. For each App description, a list of manually identified feature describing phrases are provided. (\cite{wu2021identifying}, 2021) & 50 & 619.98 & 0.30 & 0.59\\
		
		\bottomrule
		
	\end{tabular}
	\label{tbl:data}
\end{table*}

\subsubsection{Testsets Selection}

% 在上一段倒数第二句写上数据集涉及了NFR和FR
% 在formal中TTR哪个高，解释一下原因
% 在informal中LD哪个高，解释一下原因
% two common summary statistics text dataset~\cite{ferrari2017pure}

For evaluation purpose, we choose the datasets used by previous studies, which also facilitate compare the performance of ChatGPT and the methods achieving state of the arts on those datasets. Therefore, we collect the testing datasets used by recent NLP4RE studies. The study inclusion criteria include:

\begin{enumerate}
	\item The study is presented in academic literature published in major RE research publication venues;
	
	\item The study provides publicly available replication packages containing codes, datasets, and results.
\end{enumerate}

The first criterion ensures that the selected dataset is used by primary SE studies. For the venues, we refer to the NLP4RE survey~\cite{zhao2021natural} and other related reviews~\cite{wang2021, dkabrowski2022analysing, kotti2023machine}. The second criterion ensures the easy and fair performance comparison. Applying the above two criteria, we finally choose one dataset for each of four evaluation tasks shown in the figure \ref{fig:procedure}. The four datasets are separately non-functional requirements (NFR) statements, App reviews, user stories, and App descriptions. The first three datasets are in English and the last one is in Chinese, which facilitates the evaluation of ChatGPT on retrieving requirements in different languages. The requirements IR tasks performed on the datasets are multi-class classification, multi-label classification, term extraction, and feature extraction respectively. For comparison purpose, only the testset of each dataset is used to evaluating ChatGPT.

In Table \ref{tbl:data}, the first column shows tasks names that are composed of the specific IR techniques and artifact types. The second column presents the brief introduction and source of the testset selected for each task. The third and forth columns denote two commonly used summary statistics of text dataset~\cite{pustejovsky2012natural, ferrari2017pure}, i.e., the dataset size and the average number of tokens contained in each text data instance. Two descriptive measures of text content~\cite{bird2009natural} are present in the last columns in order to better understand the characteristics of the different requirements artifacts contained in each dataset. Type-token ratio (TTR) is calculated by dividing the number of unique words by the total number of words in the dataset, denoting the vocabulary diversity. Lexical density(LD) is the percentage of content words in a dataset, suggesting the content richness. On the two classification datasets, specialized statements have a much higher TTR than App reviews when expressing NFR. The NFR statements are authored by requirements analysts and developers, who should use more specific and unambiguous technical words to state the quality constraints on the software systems~\cite{ezzini2021using}. In contrast, when posting App reviews, users, mostly without technical background, tend to use more general yet ambiguous words when expressing their experience of App quality attributes~\cite{pagano2013user, wang2017aspect, jha2019mining}. In the two extraction datasets, similar TTR difference are also observed stemming from analogous reasons. The LD measures of first three datasets are all about 0.44. The App descriptions dataset has a obviously higher LD of 0.59 because App descriptions are advertisement text containing not only requirements information, e.g., highlighted features, but also diverse requirements irrelevant information, e.g., slogans, membership, and contact information~\cite{jiang2019recommending, wu2021identifying}. Those dataset and text content descriptive statistics provide better understanding of the evaluation tasks and are beneficial to investigating ChatGPT's ability to retrieve requirements information.

\subsubsection{Prompts Construction}

% peer coding
The prompts have an significant impact on the output quality thus performance of LLMs on downstream NLP tasks~\cite{liu2023pre, zhao2023survey}. To interact with ChatGPT, we handcraft the prompts for each task following the peer coding procedure~\cite{saldana2021coding} employed by previous SE research~\cite{pagano2013user, dkabrowski2022analysing}. The first two authors independently craft task prompts based on carefully reading and comprehending the papers listed in Table \ref{tbl:data}. Both of them have the background of SE research and the experience of practical software development. Three basic guidelines are followed when crafting the tasks prompts:

\begin{enumerate}
	\item Each prompt should be expressed as a single imperative or interrogative sentence;
	
	\item The words used in each prompt should come from the corresponding reference paper;
	
	\item Each prompt should describe the input artifact type, task type, and answer scope.
\end{enumerate}

% After coordination, the final 
The above guidelines ensure that each task prompt has a brief style and contains vital and necessary information. Recent studies ~\cite{wang2022super, gao2023exploring} show that detailed and longer prompts do not necessarily enhance the LLMs' performance on downstream tasks. We refine the prompts  iteratively with the training set of the selected datasets. Finally, prompts are determined as follows:

\begin{tcolorbox}[colback=gray!10]	
	(1) \textbf{NFR multi-class classification}: \textit{Tag one quality label from (Usability, Security, Operational, Performance) for the following non-functional requirement statement.}
	
	(2) \textbf{App review NFR multi-label classification}: \textit{Does the following App review involve non-functional requirements about usability (security, operational, performance).}
	
	(3) \textbf{Term extraction}: \textit{Extract all single-word and multi-word noun terms from the following software requirements statement.}
	
	(4) \textbf{Feature extraction}: \textit{Extract a list of feature describing phrases from the following Chinese App description.}
\end{tcolorbox}

For the App review multi-label classification task, we craft 4 binary relevance prompts aligning with the experimental setup in ~\cite{jha2019mining}, whose method acts as a baseline in the subsequent evaluation analysis phase.

% The first guideline ensures the prompts being brief as 
% 三个原则的目的分别是什么
% long prompts不一定好\cite{gao2023exploring}
% corresponds to the cells 

% \multicolumn{6}{p{17.8cm}}{\footnotesize Note: Average length (Avg. Len.) is the average number of words contained in each instance of a dataset. Type-token ratio (TTR) is calculated by dividing the number of unique words by the total number of words in the dataset, denoting the vocabulary diversity. Lexical density(LD) is the percentage of content words in a dataset, suggesting the content richness.}

% 过于复杂的prompt可能会降低准确性，哈工大的评测

\subsubsection{Querying ChatGPT}

We automate the querying process by calling the OpenAI API\footnote{https://platform.openai.com/docs/api-reference/chat}. Every individual instance of each testset is provided to ChatGPT in one API call. Each query is a combination of the task prompt and a requirement instance as illustrated in the right bottom part of Figure \ref{fig:procedure}. This query format and the single-turn manner jointly guarantee the zero-shot setting. Our evaluation period is from 2023-03-16 to 2023-03-28. The parameters of ChatGPT API calling are set as follows:

\begin{itemize}
	\item model: gpt-3.5-turbo. The model version we used in our evaluation is gpt-3.5-turbo, which is the latest model during our evaluation period;
	
	\item temperature: 0. This parameter determines what sampling temperature to use, between 0 and 2. Higher values like 0.8 will make the output more random, while lower values like 0.2 will make it more focused and deterministic. We set the temperature to 0 to facilitate the reproduction of our evaluation results;
	
	\item other parameters: default. Other parameters are set to default values except the model version and temperature.
\end{itemize}

% 做实验的时间是，那个时候最新的稳定模型版本是GPT-3.5-turbo
% and its default parameters are used except the parameter temperature is set to 0, which ensures the reproduction of our experimental results.

\subsubsection{Results Analysis}

ChatGPT's responses to the queries are not fully structured due to its language model nature. In our evaluation, for the classification tasks, each response is composed of a single-word label followed by some explanation sentences. For the extraction tasks, each response is an itemized list. We therefore parse the ChatGPT's responses to obtain structurally organized results. Regular expressions are employed to automatically obtain the predicted labels for the classification tasks. To compare with the ground truth, we perform lemmatization and human annotation respectively for the results of term extraction and feature extraction following the same previous practice~\cite{zhang2022automatic, wu2021identifying}.

For performance measures, we adopt the commonly used precision, recall, and $F_{\beta}$~\cite{schutze2008introduction}, which are defined as follows:

\begin{equation}
Precision = \frac{TP}{TP + FP}
\end{equation}

\begin{equation}
Recall = \frac{TP}{TP + FN}
\end{equation}

\begin{equation}
F_{\beta} = (1 + \beta^{2}) \frac{Precision \times Recall}{\beta^{2} \times Precision + Recall}
\end{equation}
where $TP$, $FP$, and $FN$ denotes the number of true positive, false positive, and false positive cases in the retrieval results. To fairly compare the performance of ChatGPT with those of baselines, $\beta$ is set to 2 for App review multi-labels classification and 1 for other three tasks, which are the same as the selected baselines.

The baselines are those methods proposed by the studies listed in column 2 of Table \ref{tbl:data}, which have achieved state-of-the-art performance on each dataset. The ground truth of the selected datasets also come from those studies. Introductions of the selected baselines are as follows:

\begin{itemize}
	\item For NFR multi-class classification, the baseline employs supervised BERT fine-tuning to build classifiers~\cite{hey2020norbert}.
	
	\item For term extraction, the baseline employs unsupervised noun phrases identification with NLP preprocessing and postprocessing~\cite{zhang2022automatic}. % pipeline
	
	\item For App review multi-label classification, the supervised SVM algorithm is used by the baseline, where multiple binary relevance sub-tasks are performed to handle the multi-label classification problem~\cite{jha2019mining}.
	
	\item For feature extraction, the baseline combines unsupervised syntactic parsing and supervised BERT fine-tuning classification~\cite{wu2021identifying}. % pipeline
\end{itemize}
% 术语抽取baseline的结果为0.77
% 我们选择这么几项工作作为baselines，因为他们提出的方法在该数据集上达到了最新的SOTA

We discuss the evaluation and comparison results in detail in the next section.

\section{Quantitative and Qualitative Analysis}

Table \ref{tbl:results} presents the performance values of ChatGPT and those of the baselines. The values of columns named ``Baselines" come from the reported results of aforementioned studies that proposed the baseline methods. For the first two tasks, the second column presents the category labels. For the other two tasks, the second column shows the software application domains. We highlight the highest average performance values with bold font for each task. In addition, we highlight the best values in the last column named ``$F_{\beta}$ Measure".

% highlight rows and the last column, f measure is a balance
% ground truth来自哪里一定要说明
% 在表注中说明feature extraction的重新计算

\begin{table*}[bt]
	\centering
	\caption{Quantitative Evaluation Results of ChatGPT on Requirements IR Tasks}
	\begin{tabular}{m{2.2cm}|c|c|c|c|c|c|c}
		\toprule
		\multirow{2}{=}{\centering \textbf{Tasks}} & \multicolumn{1}{c|}{\textbf{Categories}} & \multicolumn{2}{c|}{\textbf{Precision}} & \multicolumn{2}{c|}{\textbf{Recall}} & \multicolumn{2}{c}{\textbf{$F_{\beta}$ Measure}} \\
		\cline{3-8}
		& \textbf{/domains}& Baselines & ChatGPT & Baselines & ChatGPT & Baselines & ChatGPT \\
		
		\midrule
		
		\multirow{5}{=}{\centering Requirements statements NFR multi-class classification (S-C)} & Usability  & 0.85  & 0.83  & 0.79  & 0.87  & 0.82  & 0.83  \\ 
		& Security  & 0.92  & 0.96  & 0.92  & 0.95  & 0.91  & 0.95 \\ 
		& Operational & 0.80  & 0.78  & 0.77  & 0.75  & 0.79  & 0.74  \\ 
		& Performance  & 0.86  & 0.84  & 0.81  & 0.91  & 0.84  & 0.85  \\ 
		\cline{2-8}
		& \textbf{Average} & \textbf{0.86}  & 0.85  & 0.82  & \textbf{0.87}  & \textbf{0.84}  & \textbf{0.84}  \\
		\hline
		
		\multirow{5}{=}{\centering App review NFR multi-Label classification (G-C)} & Dependability & 0.67  & 0.48  & 0.58  & 0.78  & 0.60  & 0.69  \\ 
		& Usability & 0.56  & 0.31  & 0.45  & 0.89  & 0.47  & 0.65  \\ 
		& Performance & 0.69  & 0.06  & 0.62  & 0.94  & 0.63  & 0.24  \\ 
		& Supportability & 0.62  & 0.29  & 0.52  & 0.77  & 0.54  & 0.58  \\ 
		\cline{2-8}
		& \textbf{Average} & \textbf{0.64}  & 0.28  & 0.54  & \textbf{0.85}  & \textbf{0.56}  & 0.54  \\
		\hline
		
		\multirow{1}{=}{\centering Term extraction from user stories (S-E)}& \multirow{3}{*}{\centering Smart Home} & \multirow{3}{*}{\centering\textbf{0.72}}  & \multirow{3}{*}{\centering0.68}  & \multirow{3}{*}{\centering0.83}  & \multirow{3}{*}{\centering\textbf{0.90}}  & \multirow{3}{*}{\centering\textbf{0.77}}  & \multirow{3}{*}{\centering\textbf{0.77}}  \\[0.9cm]
		\hline
		
		%		\multirow{5}{=}{\centering App Review Multi-labels classification (coarse-grained)} & Bug Report & 0.85  & 0.56  & 0.91  & 0.84  & 0.88  & 0.67  \\ 
		%		& Feature Request & 0.86  & 0.32  & 0.83  & 0.57  & 0.85  & 0.41  \\ 
		%		& User Experience & 0.89  & 0.23  & 0.94  & 0.97  & 0.91  & 0.38  \\ 
		%		& Text Rating & 0.85  & 0.73  & 0.90  & 0.61  & 0.87  & 0.66  \\ 
		%		\cline{2-8}
		%		& Average & 0.86  & 0.46  & 0.90  & 0.75  & 0.88  & 0.53  \\
		%		\hline

		\multirow{11}{=}{\centering Feature extraction from App descriptions (G-E)} & business & 0.54  & 0.87  & 0.73  & 0.55  & 0.62  & 0.66  \\
		& tool & 0.47  & 0.86  & 0.82  & 0.72  & 0.60  & 0.78  \\ 
		& travel & 0.40  & 0.88  & 0.85  & 0.72  & 0.54  & 0.77  \\ 
		& social & 0.51  & 0.78  & 0.69  & 0.83  & 0.59  & 0.80  \\ 
		& news & 0.35  & 0.79  & 0.86  & 0.65  & 0.50  & 0.68  \\ 
		& navigation & 0.40  & 0.83  & 0.55  & 0.88  & 0.47  & 0.86  \\ 
		& music & 0.39  & 0.82  & 0.76  & 0.74  & 0.51  & 0.75  \\ 
		& life & 0.55  & 0.81  & 0.71  & 0.59  & 0.62  & 0.68  \\ 
		& education & 0.42  & 0.95  & 0.76  & 0.73  & 0.54  & 0.82  \\ 
		& entertainment & 0.29  & 0.80  & 0.77  & 0.74  & 0.42  & 0.73  \\
		\cline{2-8}
		& \textbf{Average} & 0.43  & \textbf{0.84}  & \textbf{0.75}  & 0.72  & 0.54  & \textbf{0.75}  \\	
		
		\bottomrule
		%\multicolumn{8}{l}{\footnotesize Note: the bold values indicate the best average results of the corresponding measures.}
	\end{tabular}
	\label{tbl:results}
\end{table*}

The performance of the feature extraction baseline is re-calculated based on the test results provided by~\cite{wu2021identifying}. Because ChatGPT cannot output false negative cases on feature extraction task as what the supervised classifier~\cite{wu2021identifying} dose. Therefore, the precision is computed as the percentage of correct features in the extracted ones and the recall is computed as the percentage of correctly extracted features in the ground truth. Besides, an extracted feature is treated as a true positive one only if it is exactly matched with or is a fine-grained sub-feature of a ground truth feature, which follows the evaluation practice in~\cite{wu2021identifying}. The results of NFR multi-class classification is the mean of performance on 15 projects, which aligns with the leave-one-project-out cross-validation setting used by the baseline~\cite{hey2020norbert}.

\subsection{Quantitative results analysis}

Overall, ChatGPT achieves competitive results compared with those strong baselines on all datasets when balancing precision and recall values (i.e., $F_{\beta}$). Specifically, ChatGPT largely outperforms the baseline with a margin of 0.21 on the feature extraction dataset while obtains approximative performance on other three datasets. Except the feature extraction dataset, ChatGPT acts consistently with higher recall and lower precision values with respect to the baselines. In contrast, ChatGPT exhibits much higher precision and a slightly lower recall values than the baseline in the feature extraction dataset with the margins being 0.42 and 0.03 respectively. The comparable recall values without losing much precision (except for the App review multi-labels classification dataset) partially show the effectiveness of ChatGPT in requirements information retrieval activities, where the recall is more preferred than the precision~\cite{berry2021empirical}. These overall observations suggest the potential ability of ChatGPT in retrieving useful requirements information from different types of artifacts in multiple languages.

On the two classification datasets, ChatGPT performs better than the baselines in $F_{\beta}$ on most categories except for operational and performance categories. This demonstrates that ChatGPT trained on large scale data encodes some requirements domain knowledge (i.e. the meaning of different NFR categories) as we do not provide any definition and positive or negative examples in the prompts. ChatGPT achieves much more balanced precision and recall values on the specialized dataset than it does on the App review dataset. This observation indicates that though encoding some NFR knowledge ChatGPT may includes much more false negative instances into the results under a zero-shot setting when classifying more noisy requirements artifacts~\cite{dkabrowski2022analysing}. 

On the two extraction datasets, ChatGPT obtains better or comparable $F_{\beta}$ values on various application categories. This further demonstrates ChatGPT's ability of encoding application domain knowledge besides some RE knowledge. The comparable recall values exhibit ChatGPT's potential of extractive information extraction in an end-to-end manner, which does not need a pipeline of NLP modules~\cite{wu2021identifying, zhang2022automatic}, e.g., lexical and syntactic analysis. The higher precision of ChatGPT on the feature extraction dataset is partially attributed to the fact that ChatGPT tends to extract more fine-grained features, which increases redundancy also enhances precision.

\textbf{Summary}: \textit{the quantitative results on our experimental datasets show ChatGPT's competitive ability to retrieve different categories of NFRs and functional feature-describing phrases from specialized and general requirements artifacts across various application domains}.

\subsection{Qualitative results analysis}

We conduct human content analysis~\cite{neuendorf2017content} on the outputs of ChatGPT considering the textual characteristics of requirements artifacts (last three columns in Table \ref{tbl:data}).

For the classification task, ChatGPT performs obviously better on the specialized requirements statements than on the general App reviews as shown in Table \ref{tbl:results}. The App reviews dataset has a larger average length and smaller TTR than the requirements statements dataset suggesting that user reviews contain more noisy topics~\cite{dkabrowski2022analysing} and use less category-indicating words. These noise and ambiguity of App reviews may present more challenges for ChatGPT causing its lower performance as illustrated by the following false positive and false negative cases:

\begin{tcolorbox}[colback=gray!10]
	\textbf{Review\textsubscript{1}}: \textit{Every movement causes the app to crash. Want to open a new drawing Nope, crash. Can't even access my old drawings now. It's a useless paid app until its fixed.}\\[-0.8em]

	\textbf{Response\textsubscript{1}}: \textit{Yes, the review involves non-functional requirements about performance as the app is crashing frequently and not allowing the user to access their drawings.}\\
		
	\textbf{Review\textsubscript{2}}: \textit{Actually in my area this is not very accurate. I have to rely on other apps. Annoying so fix it.}\\[-0.8em]

	\textbf{Response\textsubscript{2}}: \textit{No, this review does not involve non-functional requirements about dependability. It only mentions the functional requirement of accuracy.}
\end{tcolorbox}

The true category label of Review\textsubscript{1} is dependability. ChatGPT incorrectly classifies Review\textsubscript{1} as a performance related NFR. This can be mainly attributed to two points. The Review\textsubscript{1}'s content contains a mixture of topics on features, bug, and price. Besides, some NFR categories are vaguely-defined and very closely related~\cite{jha2019mining}, e.g., performance and dependability involved in Review\textsubscript{1}. These two reasons lead to the particularly low precision (0.06 in Table 2) of ChatGPT on classifying performance related App reviews that constitute only 1.78\% (32 out of 1800) of the total testset.

The Review\textsubscript{2} has a true label of dependability while ChatGPT predicts it as a functional issue according to the general word ``accurate". Though the word ``accurate" is identified as an indicator word through manual analysis and statistical filtering on the App reviews training dataset~\cite{jha2019mining}, ChatGPT do not use that training set in our zero-shot evaluation setting.

For the extraction task, ChatGPT achieves a better performance on the specialized requirements artifacts either as shown in Table \ref{tbl:results}. As can be seen in the last two rows of Table \ref{tbl:data}, the user stories have a lower average length and LD due to its fixed template, while they have a higher TTR stemming from the fact that they are collected from crowd workers in an encouraging creativity settings~\cite{murukannaiah2016acquiring}.

Further investigating the extracted terms by ChatGPT, we find the average length of extracted terms is equal to that of ground truth terms. However, there are many general noun phrases included into the extraction results, e.g., ``someone" and ``something similar". These observations partially explain the lower term extraction precision achieved by ChatGPT compared with the baseline which perform some task-specific filtering. The higher term extraction recall shows ChatGPT's potential NLP ability on diverse user requirements. On the feature extraction dataset, ChatGPT tends to extract fine-grained features leading to an average length of 4.35 that is greater than the average length of 2.56 for the ground truth. The following box presents some fine-grained features extracted by ChatGPT with the ground truth phrases in parentheses\footnote{We provide the translation results (Chinese to English) in the box for easily understanding purpose.}:

\begin{tcolorbox}[colback=gray!10]
	% 完全匹配
	% 具体的子特征
	
	\textbf{Education Apps}: \textit{online live learning platform (live); modify the chinese definition of a word (modify the chinese definition); bilingual/monolingual mode(bilingual mode)}\\[-0.8em]
	
	\textbf{Travel Apps}: \textit{enjoy travel discounts (enjoy discounts); share through WeChat (share)}\\[-0.8em]
	
	\textbf{Tool Apps}: \textit{clean up screenshots, similar pictures and other useless photos (clean up useless pictures); comprehensive internet security guarantee (security guarantee)}\\[-0.8em]
	
\end{tcolorbox}

The higher LD of the App descriptions dataset owes to its rich content including advertising slogan, contact and subscription information, content provided by the Apps (e.g. entertainment, news, music Apps). ChatGPT achieves a precision of 0.84, nearly twice the precision of the baseline, showing its stronger ability of discriminate between feature-related and feature-irrelevant information. 

\textbf{Summary}: \textit{the qualitative results on our experimental datasets further provide evidence for ChatGPT's quantitative performance including the powerful NLP ability and limited RE domain knowledge}.

\subsection{Threats to Validity}

% 结构化，标准化

% 正则表达式来抽取
% 温度设置为0

%The API is non-deterministic by default.  Setting temperature to 0 will make the outputs mostly deterministic, but a small amount of variability will remain.
%To mitigate this, we set clear guidelines to specify principles of crafting task prompts. We follow 
%The process peer-coding procedure 
%The authors responsible for this work major in software engineering and have a good understanding of requirements information retrieval. 
% 在制定guidelines和最终确定prompt的时候都有与后两位作者进行协商
% 输出参数的固定为0，按照网站最新说法
The major threat to internal validity derives from the prompts used to query ChatGPT. Different prompts may lead to variation of ChatGPT's responses thus influencing the performance. To craft a suitable but not perfect prompt for each task, we follow the peer-coding procedure to iterate the prompts refinement process. The first two authors are responsible for prompts construction and both of them have the background of computer science education and practice, which ensures their good understanding of requirements information retrieval. The prompts construction guidelines and the resulted prompts are also confirmed by the last three authors who have rich SE and information systems experience.

Another threat to internal validity is the non-deterministic property of ChatGPT API calling\footnote{https://platform.openai.com/docs/guides/gpt/faq}. This means that a slightly different response is returned in every call, even if the prompt stays the same. We mitigate this by two means. On the one hand, the parameter temperature is set to 0 make the outputs mostly deterministic. On the other hand, we employ parsing process to acquire structurized and normalized results from the natural language responses. For example, under the temperature of 0, ChatGPT may assign a given App review the same category label with slightly different explanations in two individual API calls, which can be also observed in the OpenAI Playground\footnote{https://platform.openai.com/playground}. With regular expressions, we parse the responses to only extract the category label, which avoid the impact of slightly different explanations on our quantitative evaluation results.

To mitigate the threats to construct validity, we choose the commonly used information retrieval measures, which are also widely used in NLP4RE studies, to evaluate the performance of ChatGPT quantitatively. 

% exhausted，本身RE领域数据集就不多
% 评论这东西emerging，且变化快，增长快，很多数据集都是不被复用的，或者说相比专业需求文档，更缺乏benchmark数据{引用综述}
The limited number of tasks and datasets in our experiments may influence the generality of our results, which constitutes the main threat to external validity. We mitigate this by selecting most common requirements retrieval tasks, i.e., classification and extraction, and public available datasets. Though there is a shortage of benchmark datasets in RE domain ~\cite{ferrari2017pure, zhao2021natural}, we select two typical research datasets for the specialized requirements artifacts. Specifically, the NFR multi-class classification dataset is built upon PROMISE Software Engineering Repository\footnote{http://promise.site.uottawa.ca/SERepository/} and the term extraction dataset is built upon Smarthome Crowd Requirements Dataset\footnote{https://crowdre.github.io/murukannaiah-smarthome-requirements-dataset/}. The volume of App Store data is massive and increases rapidly~\cite{martin2016survey, al2019app}, while the typical benchmark dataset is also relatively absent as most studies usually build datasets separately to evaluate their proposed methods~\cite{dkabrowski2022analysing}. Therefore, we select two App store datasets published by recent SE studies for general requirements artifacts.

\section{Research Implications}

In this work, we empirically evaluate ChatGPT's ability of retrieving requirements information under a zero-shot setting. The quantitative and qualitative results demonstrates ChatGPT's competitive performance compared with the strong baselines. The most significant implication of our evaluation lies to demonstrating the feasibility and promising potential of employing LLMs to perform various requirements information retrieval tasks in an end-to-end manner. 
The facility of performing multiple requirements IR tasks in an end-to-end manner with a single LLM will largely the efficiency of RE activities, especially the requirements elicitation and analysis. Recent LLMs leaderboards, e.g., AlpacaEval~\footnote{https://tatsu-lab.github.io/alpaca\_eval/} and Chatbot Arena Leaderboard~\footnote{https://huggingface.co/spaces/lmsys/chatbot-arena-leaderboard}, display that some open source models with less parameters size have achieved similar or better NLP ability to gpt-3.5-turbo. Therefore, we carefully suggest the following possible directions of future efforts on the NLP4RE research and practice based on our evaluation results and the advancement of LLMs:

\begin{enumerate}
	\item Prompt learning for specific requirements retrieval and other NLP4RE tasks could be studied to more sufficiently utilize the domain knowledge encoded by general LLMs.
	
	\item It is necessary to devise RE domain LLMs based on open source general LLMs with RE domain datasets and instructs to further improve the performance of LLMs on various NLP4RE tasks.
	
	\item To better support the above two research directions, it is also essential to build high-quality benchmark requirements datasets for training and comprehensively evaluating the devised RE domain LLMs.
	
	\item It is also valuable to investigate how to combine RE domain LLMs and the formal methods to better support requirements modeling and verification efficiently.
	
	\item RE practitioners, e.g., analysts, could further fine-tune the RE domain LLMs with their own requirements which are usually business confidentiality. The fine-tuned models can provide requirements analysis assistance with an chat-based and end-to-end manner.

\end{enumerate}

% 	, e.g., Alpaca~\cite{taori2023alpaca} more attention could be paid to devise more accurate requirements retrieval models based on open source generative LLM, e.g., Alpaca~\cite{taori2023alpaca} with .
 
% LLM exhibits the promising potential to support the development of a unified tool retrieving different types of requirements information. Furthermore, a chat-based, end-to-end tool could effectively assist requirements analysts. However, it is not appropriate to develop the tool directly using ChatGPT, as it may pose a threat to the confidentiality of requirements for those non open source software systems.

% 端到端的方式提高需求收集和分析的效率
% 因此，结合the advancement of LLMs和我们的评测，
% 因为很多评测已经表明，很多开源的能力已经达到GPT-3.5，similar NLP ability

% 从这个研究和实践角度阐明意义，
% 

\section{Related Work}

NLP techniques have long been applied to facilitate RE process to support human analysts to carry out requirements analysis~\cite{abbott1981software}. Lexical and syntactic analysis are widely used to design domain-adapted methods or tools to support retrieving various requirements information from textual artifacts, such as abstraction~\cite{peng2021environment}, terminology~\cite{zhang2022automatic}, and other requirements knowledge~\cite{lian2016mining}. The most commonly used NLP techniques in requirements IR includes tokenization, part of speech tagging, lemmatization, and syntactic parsing, which are often conducted with off-the-shelf NLP tools, e.g., Stanford CoreNLP, GATE, and NLTK. Other than those basic NLP techniques, classification~\cite{maalej2016automatic}, clustering~\cite{chen2005approach, nema2022analyzing}, summarization~\cite{tao2020identifying}, and topic modeling~\cite{tushev2022domain} are also employed to mine requirements information from large collections of textual artifacts. 

% used input in NLP4RE research.
% The most common input requirements artifacts includes 
% traditional input type
% 夏鑫 stack over flow
Software requirements specifications are traditionally the most common input document type for NLP4RE research. As the growing of online user communities in the last decade, user generated content have become prevalent in NLP4RE research, e.g., user posts and reviews in Stack Overflow~\cite{gao2020generating}, App stores~\cite{zhang2023exploring}, and social media~\cite{guzman2017exploratory}.  Feature request~\cite{jiang2019recommending}, bug report~\cite{DBLP:conf/icse/HaeringSM21}, and user opinions~\cite{lin2022opinion} are the main retrieved requirements information types. In the same time, NLP techniques also achieved rapid advancement with the help of deep learning. Word embeddings and pre-trained language models are also widely used in NLP4RE research, especially in large scale user feedback mining~\cite{gao2020generating, DBLP:conf/icse/HaeringSM21, nema2022analyzing, zhang2023exploring}.

Since the launch of ChatGPT on November 30, 2022, it has become a hot research topic to evaluate and apply ChatGPT on various domains~\cite{fraiwan2023review}, which can been glanced through the explosive growth of the volume of ChatGPT related papers. In the SE research community, ChatGPT is employed to perform various SE tasks, e.g., bug fixing~\cite{sobania2023analysis} and unit test generation~\cite{yuan2023no}. Our empirical evaluation of ChatGPT on multiple requirements IR tasks provides both quantitative and qualitative evidence for the promising potential of ChatGPT on retrieving requirements information. Therefore, our study can be used as a starting point to inspire researchers and practitioners to investigate more effective LLMs based NLP4RE methods and tools.

% used as a starting point to 
% the volume of research work on the evaluation and domain application of ChatGPT 
% explosive growth
% 总，谁做了什么，谁做了什么，总，这是最简单的模式

% 逻辑就是先说NLP4RE，说三条线，传统的，embeddings-深度学习的，再说LLM的
% 再说chat评测论文急剧增长，各个领域都是，但是illustrative examples的东西比较多
% 我们是，rather than 例子，but 定量评测，为开启设计基于LLM的需求分析工具和方法迈开了第一步

% related work 能不能放到最后说，
% LLM for software testing~\cite{yuan2023no}
% 根据这个仓库，应用几篇论文：https://github.com/saltudelft/ml4se
% bug fixing~\cite{sobania2023analysis}
% propmt由前两个作者拟定，这两个都有CS研究和工作背景，并且由后两个专家确认
% not exhausted

\section{Conclusion and Future Work}

In this work, we conduct a preliminary evaluation of ChatGPT on retrieving requirements information, specifically NFR, features, and domain terms. The quantitative and qualitative results demonstrates its promising potential. We further discuss the implications of ChatGPT for the NLP4RE research and practice based on our evaluation results.

In future, we will perform more extensive evaluations of ChatGPT and other open source LLMs on a wider range of NLP4RE tasks and datasets. Also, qualitative case studies with developers, including questionnaires and interviews, shall be conducted to better understand the practical helpfulness of LLMs in facilitating the RE process.

% Submissions are not required to reflect the precise reference formatting of the journal (use of italics, bold etc.), however it is important that all key elements of each reference are included.
\bibliography{chatref}

\end{document}